A POSSIBLE MODEL TO HIGH $T_C$ FERROMAGNETISM IN GALLIUM MANGANESE NITRIDES BASED ON RESONATION PROPERTIES OF IMPURITIES IN SEMICONDUCTORS

HIDENOBU HORI, AND YOSHIYUKI YAMAMOTO

School Materials Science, Japan Advanced Institute of Science and Technology (JAIST), 1-1 Asahidai, Nomi, Ishikawa 923-1292 Japan

SAKI SONODA

School of Engineering Science, Osaka University, 1-3 Machikaneyama, Toyonaka 560-8531, Japan

*To whom correspondence should be addressed. H.Hori (h-hori@jaist.ac.jp)

Abstract. The high $T_C$ ferromagnetism in (Ga,Mn)N were observed and almost all results are approximately similar to the experimental results in (Ga,Mn)As except the value of $T_C$. Though all standard experiments on magnetism clearly support the results, the value is unexpectedly high. This work present and discuss the possibility of high $T_C$ ferromagnetism, after brief review of the experimental results. The key speculation to Bosonization method in three dimensions is resembled with the problems in Anderson localization.

1. Introduction

   Ferromagnetism on Diluted Magnetic Semiconductors (DMS) has come to be center of attention to the researchers because of the possibilities to new functionalities in the main stream of the application to electronics. Besides the potential of such applications, it is specially noticed that the problems on the diluted impurity systems or low carrier density in semiconductors seem to provide us common interesting physical problems. For example, high efficiency in thermo-electric transformation effect in room temperature and high $T_C$ super conductivity are considered to be the typical phenomena.

The highly efficient thermo-electric transformation in BiTe[1] and high $T_C$ superconductivity materials have the carrier concentrations less than $\sim 10^{27}$ per m$^3$ ($\sim 10^{21}$ 1/cc ) and the impurity concentration of several percents. The high $T_C$ ferromagnetism is also observed in the semiconductors with these electronic conditions $10^{25}\sim 10^{28}$ per m$^3$. It is noticed that these effects are appeared in relatively high temperature range, though the impurities are randomly distributed and in low concentration. Thus, these results make us infer that some unrecognized common property may exist in the diluted impurity systems in some semiconductors.  The effects also appear under these conditions as the "unthinkable phenomena" in relatively high temperature range. In the present work, we will try to look for the physical key concept through the detailed investigation to the high $T_C$ ferromagnetism in (Ga,Mn)N.

The first observation of DMS ferromagnetism was reported for the ferromagnetism with $T_C$ of 110 K on $Ga_{1-x}Mn_xAs$ ($1 \geqq x \geqq 0$) ( the family of the Mn-doped crystals $Ga_{1-x}Mn_xAs$ is simply written by (Ga,Mn)As and similar abbreviation is also used for any other doped crystals) [2]. The value of $T_C$ = 110 K is relatively high for the Mn impurities with the concentration of 3%. After this work, a number of observations on DMS-ferromagnetism have been reported for various materials until now. Recently, our group reported the extremely high $T_C$ value of 940 K in (Ga,Mn)N[3,4].  Besides these reports, ferromagnetism of $Ca_{1-x}La_xB_6$ has been specially attracted our attention, because all ions in the crystal are not magnetic. Moreover, the estimated $T_C$ is about 600 K[5]. However, some careful experiments show the existence of the iron impurities with the concentration of about 100ppm[6] and the Auger experiment shows that thickness of the Fe impurities are deeper than 1 μm[7]. Here, the problem of "what is the origin of the ferromagnetism for the extremely low concentration of iron ions?" is emerged. Though some models have been proposed to this question, one of the most possible one is considered to be the model of DMS-ferromagnetism.

We, at first, experimentally confirm the ferromagnetism and unbelievably high $T_C$ value on (Ga,Mn)N in comparing with the possibility of super-paramagnetism. Though these experiments are quite standard ones, but some theorists have strongly disapproved the possibility of high $T_C$ value. Thus, we will point out some naive questions to their calculation based on the physical picture as the experimentalist and moreover, we want to speculate and present one possible model to explain the experimental results of the high $T_C$ ferromagnetism on (Ga,Mn)N. The important hint to the problems on high $T_C$ ferromagnetism is given by the idea to the lasing effect. The most remarkable point of the lasing effect is that the lasing action is considered as a kind of phase transition in high temperature[8]. The important point is that the lasing action is appeared in high temperature and is produced in the randomly distributed rare impurity media. In present work, referring to the conditions of lasing action, the possibility of the high TC ferromagnetism in (Ga,Mn)N is to be discussed. After that, we will briefly argue about the adequacy and applicability of the key concept used in the model.

2. The problems to ferromagnetism on DMS materials of (Ga,Mn)N.

For GaN based DMS, Dietl et al. suggested that Mn doped GaN would have high $T_C$ exceeding room temperature[3]. Recently, present authors reported successful growth of III-V based wurtzite (Ga,Mn)N films grown by Molecular Beam Epitaxy (MBE) and firstly observed their spontaneous magnetization in 0.1T, magnetization process, hysteresis. The details of magnetic properties are discussed in ref. 4 and the results are summarized in Fig. 1. The Curie temperature is roughly estimated by fitting the spontaneous magnetization curve (derived from the molecular field model using Brillouin function with $S = 5/2$) to the temperature dependence of magnetization. The steep drop of magnetization around 10 K is not included in the fitting curve. The data was taken up to 750 K as shown in Fig.1(A) and the estimated $T_C$ value is 940 K. The detailed method of sample preparation have been discussed and established now and the description of the method is described elsewhere[3,4].
These results make us feel quite anomalous high $T_C$ value and this point is central point of this work. To show the conformation of these results, the experimental procedures are reviewed in successive sub-section.

2.1. Experimental procedures and the results for (Ga,Mn)N.
The crystal structure and stacking structure of layers in the sample is shown in Fig. 2. The stacking structure is a typical one used in present work for (Ga,Mn)N film growth on a sapphire(0001) substrate: Before the growth of (Ga,Mn)N films, non-doped GaN layers were grown as the buffer layer with the thickness of 200 nm (2000 Å). Two types of buffer layers were used in present work. For one sample, the buffer layer was grown by $NH_3$-MBE method, in which $NH_3$ gas is used in MBE as a Nitrogen source. As for another method, the buffer layers were grown by rf-plasma of Nitrogen gas and the sample films are grown in the different vacuum chamber of MBE.

Mn concentrations x in $(Ga_{1-x}Mn_xN)$ of the films were estimated with the aid of an Electron Probe Micro Analyzer (EPMA). As a typical data for Mn distribution along the growth direction, Fig. 2 shows the depth profile of a sample with ~ 7% of Mn whose thickness is 2000 Å measured by EPMA. The depth dependence in Fig.2(C) measured by D-SIMS (dynamic secondary ion mass spectroscopy) and this data clearly shows uniform distribution of the Mn atoms in the whole sample space with the resolution of 20 nm. The uniform distribution along surface direction was also confirmed within the resolution of 100 nm. These mean no grain like structure in the sample space with in the resolution.

The X-ray diffraction is clearly observed and Wurtzite crystal structure is confirmed. Though the broadening diffraction spots are observed in Mn doped GaN, but the basic structures do not change. The experiments of XAFS and CAISIS are observed for the (Ga,Mn)N. The measurements of EXAFS on (Ga,Mn)N were carried out at the BL38B1 beam line of SPring-8. The experimental results give us information of substitutional doping by comparing the coherent effect of X-ray radiations from the surrounding ions around Mn and Ga. The results clearly show substitutional dope of Mn ion into Ga site of wurtzite structure[9]. Therefore, the segregation of Mn atoms or other Mn alloys are ruled out.

2.2. Magnetic measurements

As is easily understood from the data in Fig.1, the value of $T_C$ is unbelievably high. The hysteresis curve clearly indicates typical property of ferromagnetism. The remnant moment means spontaneous spin-polarization and existence of domain area. The domain size is given by the energy balance between exchange energy and summation of spin dipole energy in the domain and the usual size of the area is larger than several hundreds nanometer. This means that the ferromagnetic particles should be larger than several hundreds nanometer, if the ferromagnetism of (Ga,Mn)N were originated from assemble of ferromagnetic grains. However, the experiments of EPMA and SIMS clearly deny the grain like structure.

Figures 1(B) and (C) are explained by coexistence model of ferro- and para- (or super-para-) magnetisms4 and ferromagnetic part is about 20% of saturation moment in the sample with 8%-Mn concentration. The concentration dependence is large around 3% Mn, but the dependence is not large above the concentration of 6%. The clear ferromagnetic part is observed in the samples with the concentration above 3%-Mn. Quite recently, we observed the steep decrease of remnant moment at zero field with decreasing temperature below 10 K, while the remnant moment is approximately constant from 10 K to 400 K[10]. Such a phenomenon is quite anomalous for the research field of magnetism.

2.3. Optical, transport and magnetic properties.

Because the ferromagnetism in (Ga,Mn)N is originated from the properties of DMS materials, investigations of transport and magnetic properties make us expect the important hints to the origin of magnetism. Optical measurements also give us useful information to the electronic states. Though the 6% Mn impurities make large change in the spectra and any peak originated from sharp and strong emission spectrum in pure GaN with the concentration of 94% is not observed. This means that the large change in electronic states of the whole host crystal and clearly deny the model of mixed crystal state, which consists of pure GaN (94 at%) and some segregated Mn compound (6 at%).

Takeyama et al. observed the hydrogenic exciton absorption spectra superimposed on the band absorption edge in the ultraviolet range[11] and it means existence of the hydrogen like atomic exciton in (Ga,Mn)N.

As (Ga,Mn)N is a DMS materials, the origin of the ferromagnetism is closely related to the properties of the carriers. In fact, Fig.4 (C) and (D) clearly show this relation. The temperature dependence to the carrier density in Fig.4 was obtained from the data of Hall resistance. It is noticed that for the Mn-concentration higher than 3% in (Ga,Mn)N, the carrier is p-type one, while pure GaN grown by $NH_3$-MBE method has n-type conduction. The GaN film has n-type carrier with the density of $10^{27}$ $m^{-3}$ ($10^{21}$ $cm^{-3}$), but in the case of a sample with Mn-concentration of 6.8%, for example, the carrier of (Ga,Mn)N has hole carrier density is $10^{26}$ $m^{-3}$ ($10^{20}$ $cm^{-3}$ ).

The GaN produced by $NH_3$-method has electron conduction and metallic temperature characteristics. A most possible model for the electron conduction in the GaN is a given by some lattice defect such as anti-site structure. As the concentration of Mn impurities increase, the carrier type changes from n-type to p-type and the ferromagnetism in (Ga,Mn)N is appeared above the Mn concentration of 3%. Though the hole density becomes higher with increasing Mn concentration, but the ratio of the ferromagnetic part to the total magnetization does not increase so much.

From the Hall resistance data, the temperature dependence on the carrier density can be reviewed as Fig. 5. As is seen in Fig.5, the increase of resistance in low temperature strongly depends on the decrease of carrier density. The hole conduction can be explained by the electron hopping motion from $Mn^{2+}$ to $Mn^{3+}$. The existence of the mixed valence state of Mn ions in (Ga,Mn)N is recently reported by our group[10]. From both data of Fig.5, the main reason of the increase in resistivity is considered to be due to the decrease of carrier density. Furthermore, the decrease of carrier concentration also coincide with the decrease of spontaneous magnetization, as is seen in Fig.4(D).

As is discussed in ref. 4, the double exchange mechanism explains the experimental results of the ferromagnetism in (Ga,Mn)N. The double exchange model is firstly proposed by the present authors in the previous work[4] and recently, the model is also theoretically discussed and supported. Electron hopping conduction between Mn atoms decrease with decreasing temperature. Therefore, the spontaneous magnetization arising from double exchange mechanism may be decreased at low temperatures. The dotted line arrow in fig 4 (c) and (d) indicates that the the threshold temperature below which the carrier trapping becomes markedly large correspond to the temperature below which the spontaneous magnetization shows the reduction.

From the experimental results, following physical picture is described: some $Mn^{3+}$ ions in (Ga,Mn)N compensate the electrons in GaN and $Mn^{2+}$ is produced near the Fermi level. Such $Mn^{2+}$ in (Ga,Mn)N splits into two virtual bound states. These virtual bound states have two opposite spin polarization for the spin on $Mn^{2+}$. The spin polarizations are produced by s-d Hamiltonian. The virtual bound states make the high density of states near the Fermi level. The two density of states have opposite spin polarizations near the Fermi level. The physical picture of the double exchange mechanism is described as the collective motion of the electron hopping in the spin-polarized band.

This mechanism can explain the drastic change in carrier type in (Ga,Mn)N at the critical concentration about 3% and this model requires coexistence of $Mn^{2+}$ and $Mn^{3+}$ ions. In fact, the experimental result of EXAFS supports this model[10]. In fact, the XANES spectra are consistent with neither $Mn^{2+}$ nor $Mn^{3+}$ and the mixed valence state is suggested[9].

2.4. Review of the results and discussion.

The magnetization measurements shows typical ferromagnetism and the possibility of segregation model is experimentally denied by hysteresis curve, existence of zero field magnetization and its temperature dependence and uniform distribution of Mn ions in the whole crystal region. We also experimentally deny the segregation model by the observation of ferromagnetic powder sample. The field dependence of step like magnetization is realized by the powder sample with the concentration higher than 90 %. Even in such a high density ferromagnetic powder sample, the hysteresis is not observed at room temperature. These phenomena clearly show that the segregated super paramagnetism with the 6% Mn impurities can not reproduce the magnetization data in (Ga,Mn)N.

Mixed valence states of $Mn^{2+}$ and $Mn^{3+}$ and double exchange mechanism based on the hopping conduction is supported by experiments of EXAFS and XANES spectra, transport property and strong coincidence in the anomalous region between transport and magnetic properties.

Especially, it is quite noticeable that the zero-field spin polarization in High $T_C$ exceeding room temperature is sharply decrease below 10 K. This effect directly means that the conduction electrons produce the ferromagnetism, because the temperature range is clearly consistent with decreasing region of the hole carriers. Thus, the problem is "what is the origin of the High $T_C$ ferromagnetism in (Ga, Mn)N with the several % concentration of Mn impurities".

3. A possible model and the application to other problems.

3.1. Question to the theoretical calculation.

Though ferromagnetism in (Ga,Mn)N is experimentally proved, some theorists clearly deny the experiments based on their calculation[9]. According to the theory, the band structure of (Ga,Mn)N is calculated by so-called KKR-CPA method. The calculation method to the band structure is a kind of APW-methods and the crystal space are separated into free electron- and atomic sphere- regions (or muffin-tin type potential region). In this method, the atomic potential region is restricted within the sphere having characteristic radius R. The radius is included in the Wigner-Seits cell.

For such procedure, we have a question that the segmentation into the Wigner-Seitz cells might be inadequate, because the imperfect shielding length in low carrier density on DMS materials is quite longer than the metals. In the case of metals, there exist enough electrons and the size of imperfect electron shielding around the cation is short enough in comparing with the size of Wigner-Seits cell and the segmentation method is quite adequate. The imperfect shielding length in (Ga,Mn)N is simply estimated by use of Thomas-Fermi approximation[12]. According to the method, the shielding length rshield is estimated from the carrier density and other basic constants. The role of atomic potential is relatively important in the imperfect shielding region. In fact, the hydrogenic exciton spectra in (Ga,Mn)N were already observed in ultraviolet region[11]. Appling the formula to the case of (Ga,Mn)N, the estimated value of the shielding size (called "Thomas-Fermi diameter" in this work) is longer than 7 nm. Thus, the size is considered to be much longer than the longest lattice constant of c = 0.517 nm. For example, the Muffin-tin radii are 0.1016 nm for cations and 0.9252 nm for anions in the theory[13]. Though the free electron region is restricted as narrow as possible in APW method and the smooth connection between core and free electron wave functions, the spin state is fairly restricted by the properties of plane wave functions. The free electron part exhibits the spin singlet state for every energy level. Such a singlet spin state is given by the resultant contribution of Coulomb- and exchange- integrals based on the plane wave functions. Thus, the Thomas-Fermi effect might give us some severe problems to the numerical estimation. Because the long range effect in the Coulomb interaction is more important in the Thomas-Fermi region, it can be considered that all magnetic ions of (Ga,Mn)N in the Thomas-Fermi region directly can contribute to the magnetic order.

3.2. A model to high $T_C$ ferromagnetism in (Ga,Mn)N.

Though ferromagnetism in (Ga,Mn)N is experimentally proved, but the remaining question is "why such high $T_C$ is realized in the conditions of diluted Mn ions, the random distribution and low carrier density". In our model, the story of the ferromagnetic order is given by the double exchange mechanism based on the virtual bound states coupled by spin correlated electron hopping. The spin correlated transfer is generated by the s-d Hamiltonian. This model is schematically given by Fig.6.

The condition of low carrier density arises from the experimental result in Fig. 4(D). These conditions are quite disadvantage for the magnetic order in usual materials. As is noticed in previous section, the important hint to the problems on high $T_C$ ferromagnetism is suggested by the idea to the lasing effect. The most remarkable point is that the lasing action is considered as a kind of phase transition in high temperature[8]. The important point is that the lasing action can work in high temperature, though the lasing media has the conditions of randomly distributed and low concentration lasing atoms. In present work, referring to the conditions of lasing action, the possibility of the high $T_C$ ferromagnetism in (Ga,Mn)N is to be discussed.

At first, we consider the one-dimensional case for the simplicity and later we will discuss about the three-dimensional case. It is noticed that the Mn impurities can be scattering center to the conduction electron and the ions on substitutional sites can form a resonator to the conduction electrons, because some electron waves can form the standing wave in the space between two Mn ions. Thus, if we assume that the role of laser light corresponds to the role of electron wave in the conduction band and a couple of Mn impurities are considered as a resonator in one-dimension. Here, we temporally call the resonator made from Mn impurities "impurity resonator" in this work. The impurity resonator to the electron waves corresponds to the resonator in laser action. On this background, we propose a lasing like mechanism to high $T_C$ ferromagnetism in (Ga,Mn)N as a possible model by using following corresponding:

Looking at the correspondence to the conditions of both Laser and High $T_C$ ferromagnetism, we can understand the similarity of them, except the statistics in 11) in the Table I. The condition 6) is satisfied as followings. At first, we want to emphasize that this work is based on following experimental results on (Ga,Mn)N. a) Existence of mixed ion states of $Mn^{2+}$ and $Mn^{3+}$ together with hole conduction in the ground state (Concentration dependence shows that there exist both n- and p- conduction types, but p-type is majority in ferromagnetic state with Mn- concentration higher than 3%, as was discussed in ref. 4. b) GaN crystals without Mn ions made by ammonia-method has n-type conductivity. c) Appearance of ferromagnetism is strongly related with conductivity as is given in the text and ref. 4. These characteristics are originated from the properties as the ground states in this crystal. For these results in (Ga,Mn)N, we may assume that the hole conduction is produced by hopping conduction between $Mn^{2+}$ and $Mn^{3+}$ states. This model just follows the "double exchange mechanism". This directly means the ground ferromagnetic state coupled with the hopping conduction. If the ferromagnetism is represented by the electron states in conduction band and two Mn ion states (mixed valence state) on Ga-sites of GaN, the ferromagnetic state looks like some excited state. In the case of (Ga,Mn)N, a lot of holes of $Mn^{2+}$ already exists as $Mn^{3+}$ ions (which is the excited state generated by the ionization of $Mn^{2+}$). The population inversion of holes is realized in the ground state when the population of $Mn^{3+}$ is higher than that of holes in virtual bound states of $Mn^{2+}$, which is across the Fermi level. Observation of the hole conduction supports existence of such mechanism. The structure is produced in the process of crystal growth of (Ga,Mn)N.

The condition 11) is possible to satisfy in the case of one dimensional case, because the scattering phenomena near $\varepsilon_F$ in the electron system can be describes by Bosonization representation. The basis of the Bosonization representation is given by standing wave states and the standing wave representation is produced by Bogoliubov transformation to the free electron states and the transformed creation annihilation operators satisfy the Bose commutation relations. This means that the one-dimensional system satisfies the condition of lasing action. In the case of High $T_C$ ferromagnetism, the "lasing" electron is standing wave state and is considered to be a kind of pairing state.

On these backgrounds, we may speculate following model:

1) In the 3-dimensional case, all electrons cannot completely describe the Bosonization representation, but for the special electron modes may satisfy the condition. This possibility is similar to the problem in the Anderson localization in highly condensed electron system. If this speculation is true, the coexistence of ferro- and para- (super para-) magnetisms becomes quite intrinsic problem and the ratio of about 20% corresponds to the ratio of Boson state modes to the total states.

2) The electronic states are given by double exchange electronic states coupled with the lasing standing waves formed in the conduction bands.

3) Origin of high temperature working temperature is given by following physical picture: because of low carrier density, the electron shielding is deficient and the Thomas-Fermi shielding length is quite long by about 7 nm and Coulomb interaction of the Mn ion and localized nature of electron becomes relatively important. Such a structure just is followed by the strong spin correlated interaction. The stability of inter Mn ion coupling depends on the stability of the standing waves in the conduction band.

4) The physical picture of High $T_C$ is explained by the stability of standing waves against the

thermal fluctuation of the lattice. The point of the stability is in the long wavelength standing waves in the long distant impurity resonator. The long distant resonator in (Ga,Mn)N can be several ten times larger than the lattice constant. Such a resonator generally keeps high efficiency as a resonator because the effect of lattice fluctuation to the resonator is much lower than the resonator with the distance of lattice constant. Thus stored standing waves with long wavelengths are relatively stable in comparing with the band edge electron waves, which make large effect to the usual metallic ferromagnetism.

Thus, present model is considered to be a possible model to explain the coexistence of Ferro- and Para- magnetisms and high $T_C$ value of (Ga,Mn)N.

3.3. Discussion

This model has a possibility to apply other phenomena produced by substitutionally doped mixed valence ions with low carrier density in semiconductors. One of the characteristic properties in the phenomena is the high working temperature. The characteristics are supported by some mode in the "impurity resonator" structure. In the three-dimensional crystal, a part of conduction mode is responsible up to high temperature. These effects should reflect to the electronic state. Some papers have been reported existence of narrow spectra near $\varepsilon_F$ even in high temperature for working range. In some X-ray photo-Emission research, such spectra are called coherent spectrum. The existence of coherent spectra have been reported in DMS ferromagnetic materials[14,15]. High $T_C$ superconductor[16] and other spin correlated materials[17]. The existence of these sharp bands in high temperature seems to support the model in present work. But the model is on the speculations that the existence of "impurity resonator" and the ratio of the standing wave in three dimensions. Present authors hope theoretical support for these problems.

ACKNOWLEDGEMENTS
The synchrotron radiation experiments were performed at SPring-8 with the approval of Japan Synchrotron Research Institute (JASRI) as Nanotechnology Support Project of the Ministry of Education, Culture, Sports, Science and Technology.

**Table I.** Corresponding elements between Lasing Action and High $T_C$ (Ga,Mn)N

|    | Laser | High $T_C$ |
|----|-------|------------|
| 1) | Lasing light | Electron wave in conduction band |
| 2) | Excited state $E_2$ | Electron trapped Acceptor of $Mn^{2+}$ |
| 3) | Lower state $E_1$ | Lower state on Ga site of $Mn^{3+}$ |
| 4) | Induced optical transition: $E_2 \rightarrow E_1$ | Electron emission : $Mn^{2+} \rightarrow Mn^{3+}$ |
| 5) | Photon absorption process : $E_1 \rightarrow E_2$ | Electron absorption: $Mn^{3+} \rightarrow Mn^{2+}$ |
| 6) | Inversion population | Hole dominated : existence of $Mn^{2+}$ |
| 7) | Induced optical transition | Quantum transitions: electron emission |
| 8) | Ratio of responsive atoms ~ 1% | Concentration of Mn impurities 3 ~ 7% |
| 9) | Electric dipole transition | Transition by s-d interaction |
| 10) | Resonator | "Impurity resonator" |
| 11) | Bose statistics of light | Fermi statistics of electron wave |

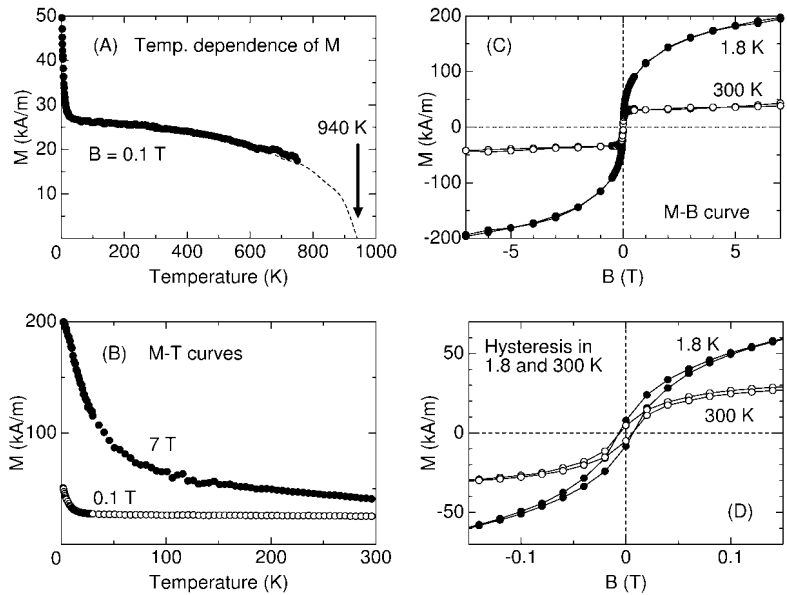

**Figure 1.** Measurements of Ferromagnetism in (Ga,Mn)N. **(A)** Temperature dependence of magnetization of (Ga,Mn)N (M-T curve) in 0.1 Tesla(T) observed up to 750 K to estimate the value of $T_C$. The value of $T_C$ is about 940 K on the theoretical curve obtained from the mean field theory. **(B)** M-T curves in low (0.1 T) and high fields(7 T). **(C)** Field dependence of magnetization M-B curve. **(D)** Hysteresis curves of the ferromagnetism in liquid He and room temperatures.

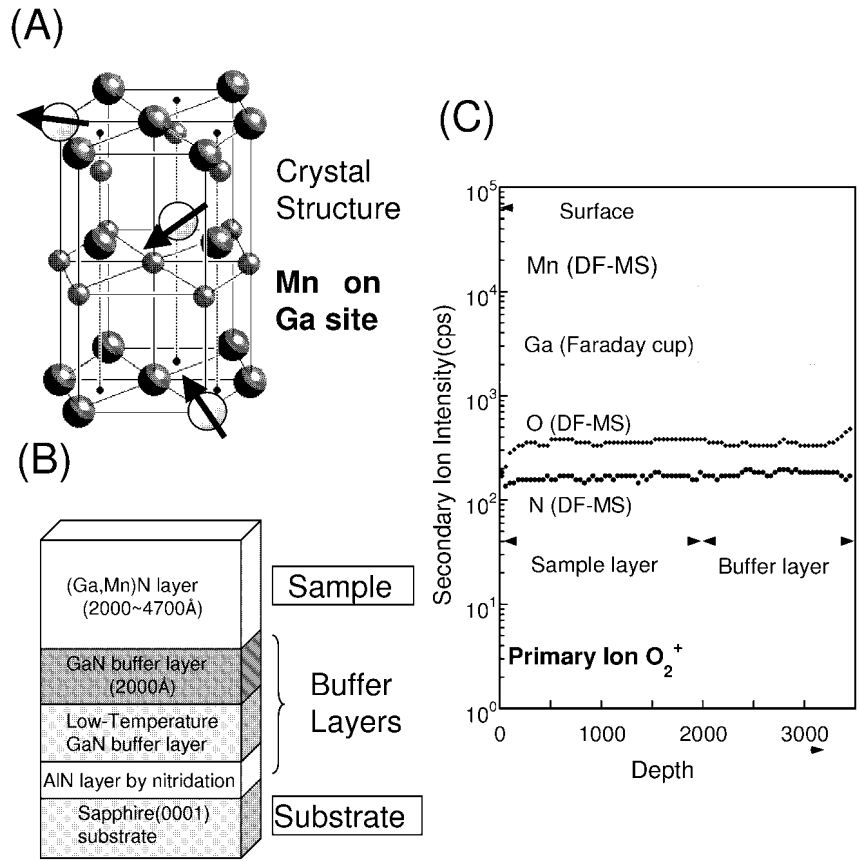

**Figure 2.** Structure of (Ga,Mn)N sample. **(A)** Crystal structure. Arrows mean spin moments on Mn impurities. **(B)** Stacking Layer structure of the sample. The top layer is (Ga,Mn)N. **(C)** Distribution of constituent atoms observed by D-SIMS.

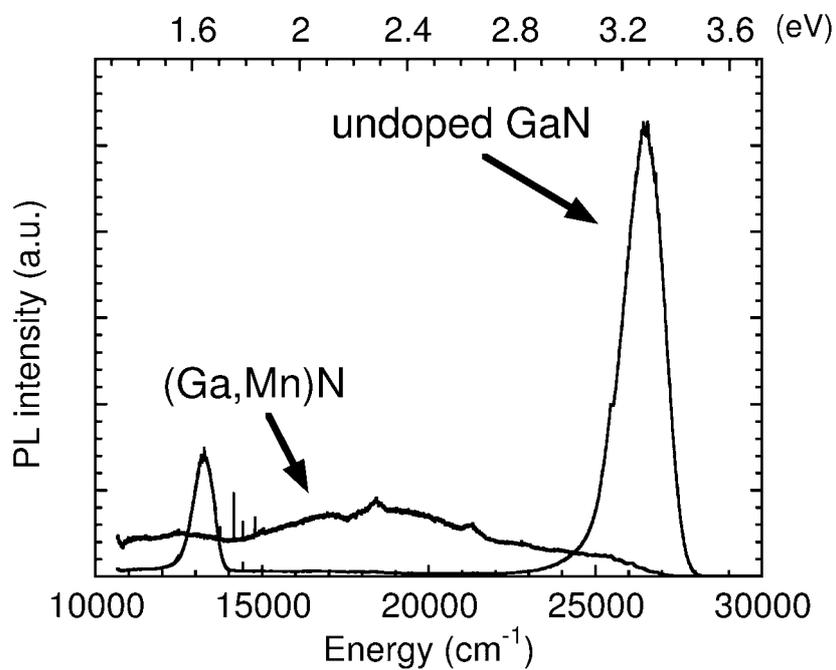

**Figure 3.** Emission spectra of GaN and (Ga,Mn)N with 6% Mn. The sharp peak near 13000 cm$^{-1}$ is considered to be second reflection by the grating in the spectrometer corresponding to strong spectrum at 26500 cm$^{-1}$. No sharp spectrum corresponding to these sharp lines in GaN is not found in the spectra in (Ga,Mn)N . This means large change in electronic state by Mn doping.

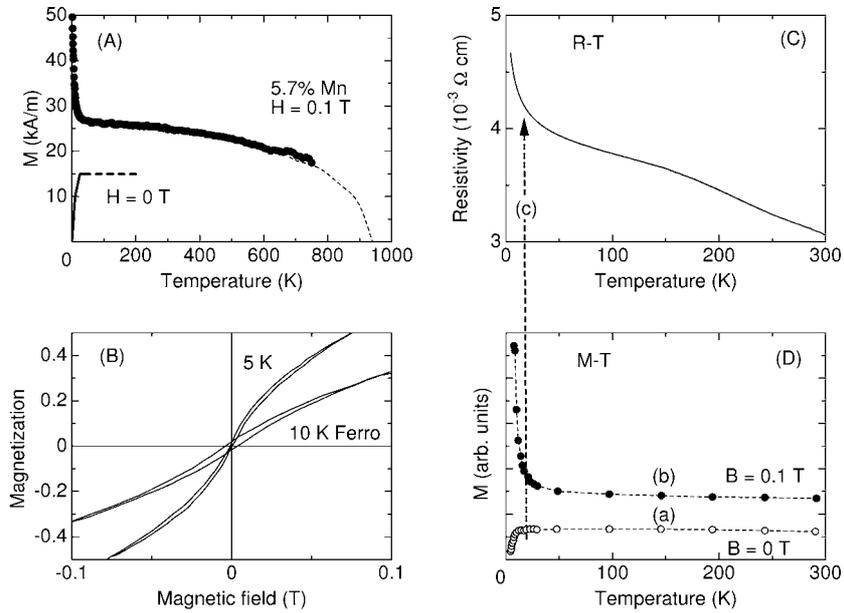

**Figure 4.** Review of M-T curves and the relation with R-T curve. **(A)** Temperature dependence of magnetization in 0.1 T and schematic curve of temperature dependence of zero field magnetization of the sample with Mn-concentrations of 8 %. **(B)** Hysteresis curves. It is noticed that coercive field of 10 K is much larger than that of 5 K. **(C)** Temperature dependence of resistance (R-T curve). The temperature at (c) is the threshold temperature below which the carrier trapping becomes markedly large. **(D)** The remnant magnetization at zero field (a) and magnetization at the field just above the hysteresis loop (b).

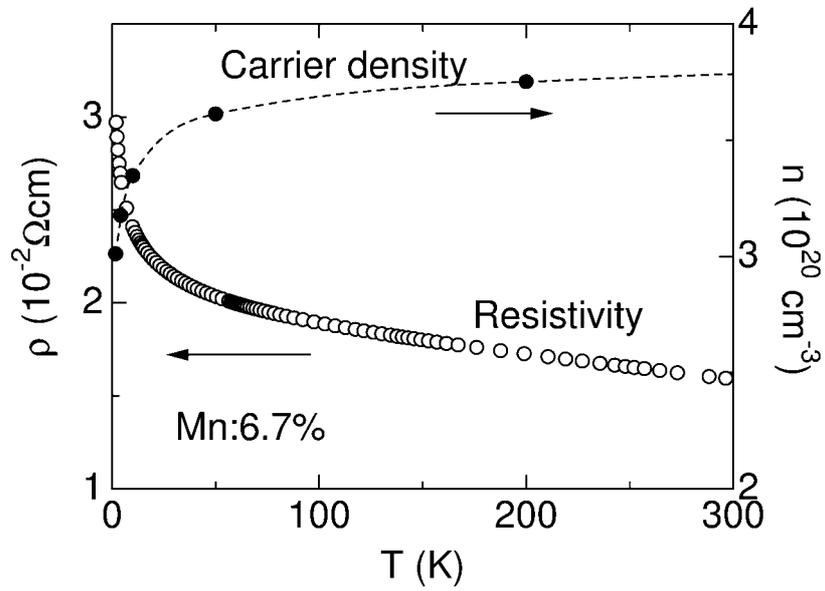

**Figure 5.** Temperature dependence of carrier density and resistivity on (Ga,Mn)N. Because of the clear correlation to the carrier density, the main reason of increase in resistivity is considered to be due to the decrease of carrier density and furthermore, the decrease of magnetization also correlated to the decrease of spontaneous magnetic polarization.

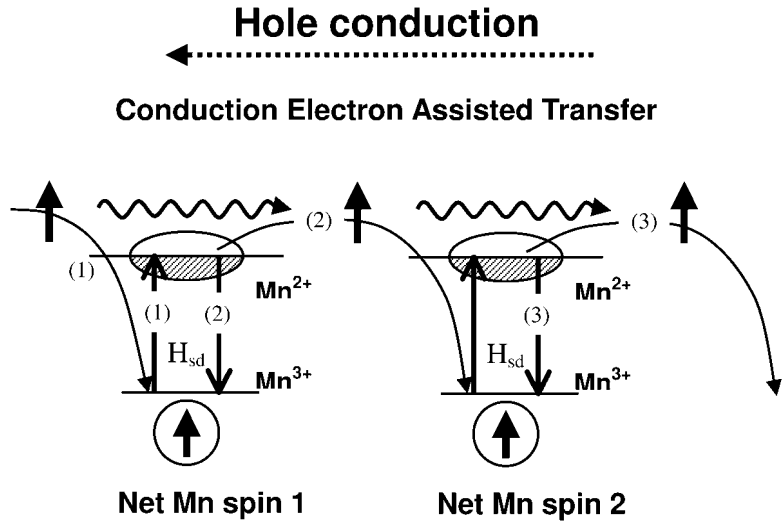

**Figure 6.** Schematic view of Double Exchange Model. Only one virtual bound state near the Fermi level is shown in this figure. Process(1) is electron trapping and processes (2) and (3) are spin correlated transfer process. A electron is trapped to $Mn^{3+}$ and the $Mn^{3+}$ ion changes to $Mn^{2+}$ near the Fermi level. The $Mn^{2+}$ is in virtual bound state and the spin state is restricted to the net Mn spins because the transition is generated by spin-correlated interaction of s-d Hamiltonian. This is the elementally process of electron hopping.